\documentclass[]{spie}  
\usepackage[]{graphicx}
\usepackage[]{subfigure}
\usepackage[]{amssymb}
\usepackage[]{mathptmx}
\usepackage[]{amsmath}
\usepackage[]{lscape}
\include{amsmath}
\title{Antenna-coupled TES bolometers for the Keck Array, Spider, and Polar-1} 

\author{R. O'Brient,\supit{a,b}
 P. A. R. Ade,\supit{c}
Z. Ahmed,\supit{k,l}
R.W. Aikin,\supit{a}
M. Amiri,\supit{d}
S. Benton,\supit{e}
C. Bischoff,\supit{f}
J.J. Bock,\supit{a,b}
J. A. Bonetti,\supit{b}
J. A. Brevik,\supit{a}
B. Burger,\supit{d}
G. Davis,\supit{d}
P. Day,\supit{b}
C.D. Dowell,\supit{a,b}
L. Duband,\supit{g}
J. P. Filippini,\supit{a}
S. Fliescher,\supit{h}
S.R. Golwala,\supit{a}
J. Grayson,\supit{k,l}
M. Halpern,\supit{d}
M. Hasselfield,\supit{d}
G. Hilton,\supit{i}
V.V. Hristov,\supit{d}
H. Hui,\supit{a}
K. Irwin,\supit{i}
S. Kernasovskiy,\supit{k,l}
J. M. Kovac,\supit{f}
C. L. Kuo,\supit{k,l}
E. Leitch,\supit{j}
M. Lueker,\supit{a}
Megerian, K,\supit{b}
L. Moncelsi,\supit{a}
C.B. Netterfield,\supit{e}
H. T. Nguyen,\supit{a,b}
R. W. Ogburn IV,\supit{a,b}
C. L. Pryke,\supit{j}
C. Reintsema,\supit{i}
J.E. Ruhl,\supit{m}
M.C. Runyan,\supit{a,b}
R. Schwarz,\supit{h}
C. D. Sheehy,\supit{h}
Z. Staniszewski,\supit{a,b}
R. Sudiwala,\supit{c}
G. Teply,\supit{a}
J. E. Tolan,\supit{k,l}
A. D. Turner,\supit{b}
R.S. Tucker,\supit{a}
A. Vieregg,\supit{f}
D. V. Wiebe,\supit{d}
P. Wilson,\supit{b}
C. L. Wong,\supit{f}
W.L.K. Wu,\supit{k,l}
K.W. Yoon\supit{k,l}
\skiplinehalf
\supit{a}California Institute of Technology, 1200 E. California Blvd., Pasadena, CA 91125 USA; \\
\supit{b}Jet Propulsion Laboratory, 4800 Oak Grove Dr., Pasadena, CA 91109, USA; \\
\supit{c}Dept. of Physics and Astronomy, University of Wales, Cardiff, CF24 3YB, Wales, UK; \\
\supit{d}Department of Physics and Astronomy, University of British Columbia, 6224 Agricultural Road, Vancouver, BC V6T1Z1, Canada; \\
\supit{e}Department of Physics, University of Toronto, Toronto, ON M5S 1A7, Canada; \\
\supit{f}Harvard-Smithsonian Center for Astrophysics, 60 Garden Street, Cambridge, MA 02138; \\
\supit{g}Service des Basses Temperatures, DRFMC, CEA-Grenoble, 17 rue des Martyrs, 38054 Grenoble Cedex 9, France; \\
\supit{h}School of Physics and Astronomy, University of Minnesota, 116 Church Street S.E.,Minneapolis, MN 55455; \\
\supit{i}NIST Quantum Devices Group, 325 Broadway, Boulder, CO 80305, USA; \\
\supit{j}University of Chicago, KICP, 933 E. 56th St., Chicago, IL 60637 USA; \\
\supit{k}Stanford University, 382 Via Pueblo Mall, Stanford, CA 94305, USA; \\
\supit{l}Kavli Institute for Particle Astrophysics and Cosmology (KIPAC), Sand Hill Road 2575, Menlo Park, CA 94025, USA; \\
\supit{m}Physics Department, Case Western Reserve University, Cleveland, OH 44106 USA; \\
}

\authorinfo{Further author information: (Send correspondence to Roger O'Brient).  E-mail: rogero@caltech.edu}

\begin{document} 
  \maketitle 

\begin{abstract}
Between the BICEP2 and Keck Array experiments, we have deployed over 1500 dual polarized antenna coupled bolometers to map the Cosmic Microwave Background's polarization.  We have been able to rapidly deploy these detectors because they are completely planar with an integrated phased-array antenna.  Through our experience in these experiments, we have learned of several challenges with this technology- specifically the beam synthesis in the antenna- and in this paper we report on how we have modified our designs to mitigate these challenges.  In particular, we discus differential steering errors between the polarization pairs' beam centroids due to microstrip cross talk and gradients of penetration depth in the niobium thin films of our millimeter wave circuits.  We also discuss how we have suppressed side lobe response with a  Gaussian taper of our antenna illumination pattern.  These improvements will be used in Spider, Polar-1, and this season's retrofit of Keck Array.
\end{abstract}

\section{INTRODUCTION}
\label{sec:intro}

The Cosmic Microwave Background's (CMB) temperature anisotropies and curl-free E-mode polarization anisotropies, both generated by scalar inflationary potentials, have been mapped by numerous experiments and used to constrain a multitude of cosmological parameters.  To date, divergence-free B-modes in the CMB, which would have been seeded by inflationary tensor potentials, have not been detected.  However, current data favor a scalar index $n_s<1$, which suggests that the tensor-to-scalar ratio $r$ may be non-zero \cite{WMAP_seven_year}.  A B-mode detection at degree angular scales would provide strong confirmation of inflation, but even upper limits on $r$ will constrain the energy scale of inflation, thus making it a fundamental parameter of interest.  Additionally, gravitational lensing can shear E-modes into B-modes, creating a smaller angular scale non-primordial anisotropy.  The details of this peak can constrain the summed mass of the different neutrino species and potentially inform us of the existence of sterile neutrinos\cite{Dodelson_white_paper}.

Our team has developed a low cost camera design with just enough resolution to detect B-modes from primordial tensor perturbations.  This design builds off the success of BICEP-1, using refracting optics with a 30 arcmin resolution\cite{Yoon_B1_SPIE}\cite{Aiken_SPIE}.  BICEP2 and Keck Array's five BICEP2 style cameras are both currently collecting data at the South Pole with nearly 1500 pixels (3000 detectors) collectively \cite{Keck_array_SPIE}.  The balloon-borne Spider will fly in 2013 from the McMurdo station in Antarctica with six cameras and a comparable pixel count\cite{Filippini_SPIE_SPIDER}.  Finally, we will deploy the 1.5m crossed-Dragone telescope Polar-1 to the South Pole in 2013 with 1900 pixels.  Its 5 arcmin resolution will be sufficient to detect lensed B-modes.

All four of these experiments use phased-array antenna-coupled TES bolometers developed at Caltech and JPL.  Our detectors are entirely planar, and hence scalable to large monolithic arrays.  This has provided our team with a crucial advantage over competing teams, allowing early deployment of high sensitivity focal planes.  As a result, our technology has matured to a point where we have identified several failure modes from both lab and field data and we have refined our original design with these measurements in mind.  This paper describes some of the efforts to improve our detectors.  The second section summarizes our detector design, the third describes how we corrected differential pointing between polarization pairs, and the last outlines how we have suppressed sidelobe response with a tapered illumination pattern in anticipation of Polar.

\section{DETECTOR AND CAMERA ARCHITECTURE}
\begin{figure}[ht]
\centering
\subfigure[BICEP2 optics \label{B2_rays}]{
\includegraphics[scale=0.5]{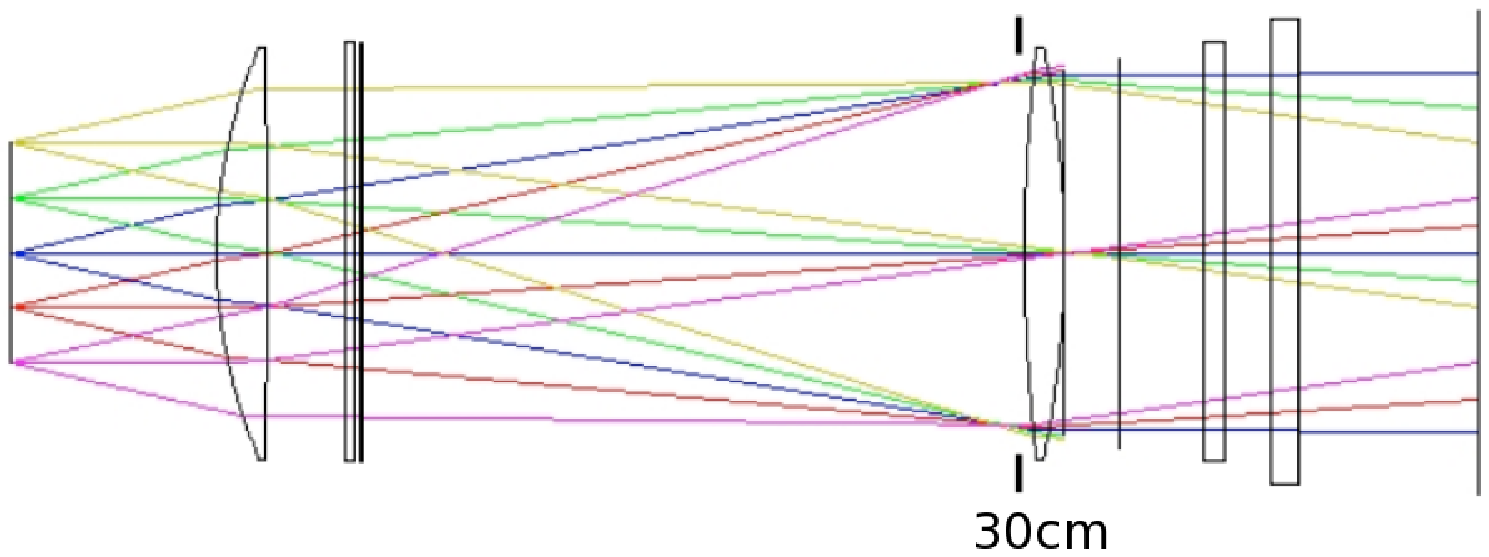}
}
\subfigure[Polar-1 optics\label{P1_rays}]{
  \includegraphics[scale=0.5]{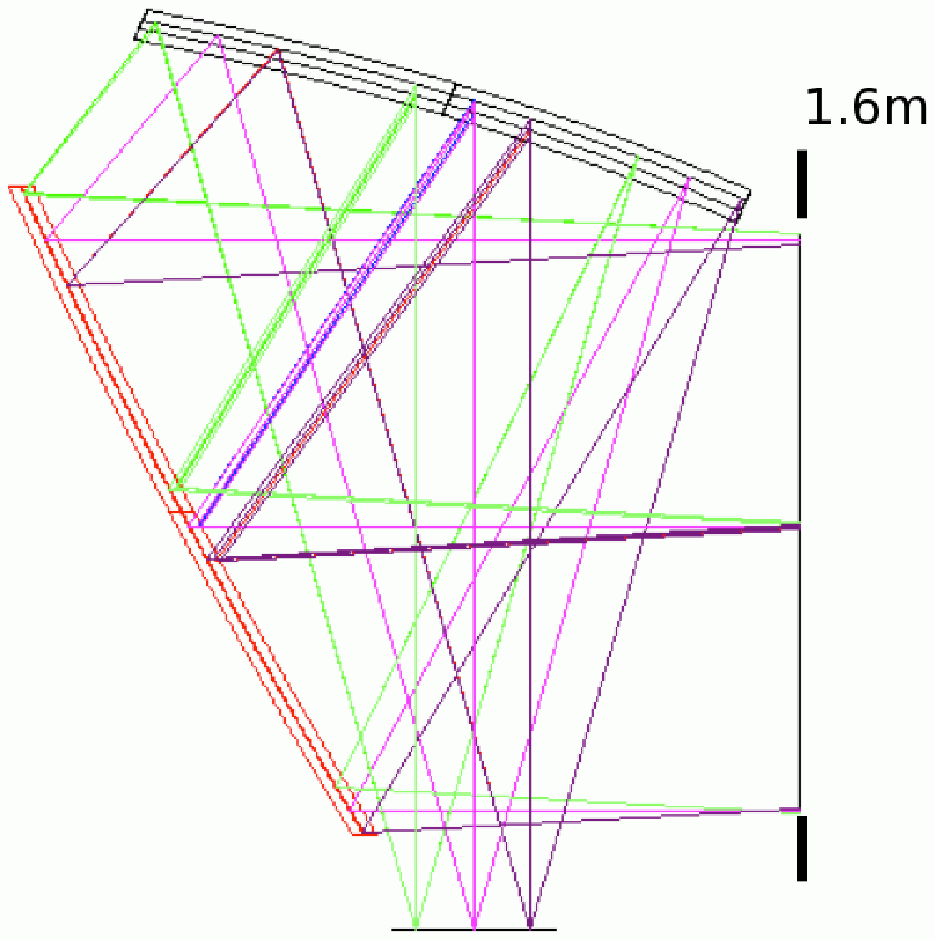}
}
\caption{Ray Diagrams for the optics in the two types of experiment.  The focal plane is at left/bottom while the aperture is at the far right of both.  Notice that the optics in BICEP2 (Figure \ref{B2_rays}) form an aperture stop that can naturally be made cold \cite{Aiken_SPIE} while those for Polar-1 (Figure \ref{P1_rays}) form a stop where cooling is more challenging.\label{optics_design}}
\end{figure}

Figure \ref{B2_rays} shows the BICEP2 optics design, which we used in Keck Array and Spider with only minor modifications.  Each camera is a refracting telescope with two HDPE lenses that image the sky onto a focal plane while forming an aperture just on the sky-side of the objective lens.  We place a cold stop at this point to terminate side lobes of the antenna coupled bolometers in our focal plane.  We also use a series of Teflon, nylon, and metal-mesh IR blockers to filter away higher frequency out-of-band power\cite{Aiken_SPIE}.

\begin{figure}
 \begin{center}
 \subfigure[]{
 	 \includegraphics[height=5cm]{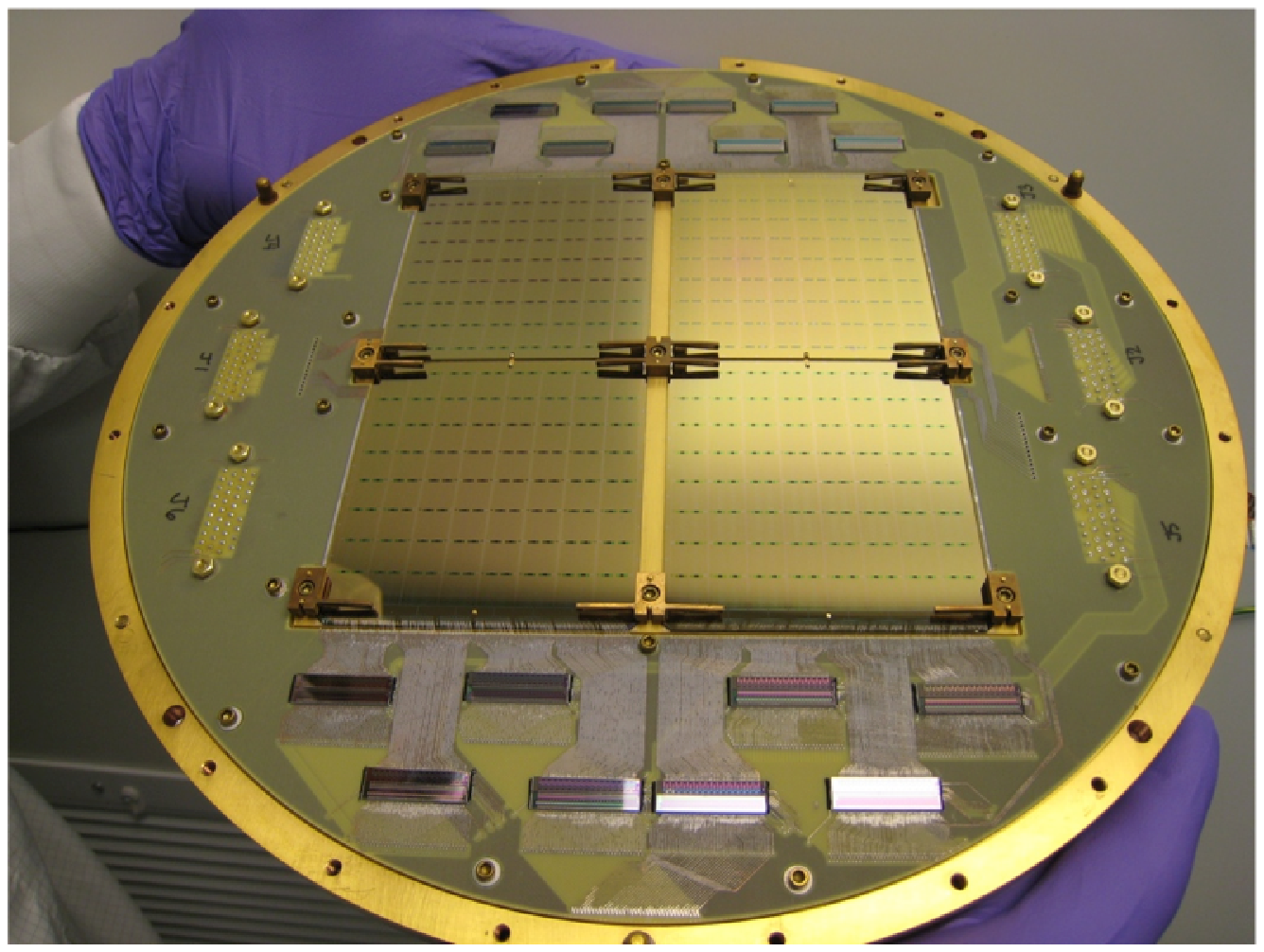}
	 	\label{FPU_pic}
	}
\hspace{5mm}
\subfigure[]{
	\includegraphics[height=5.6cm]{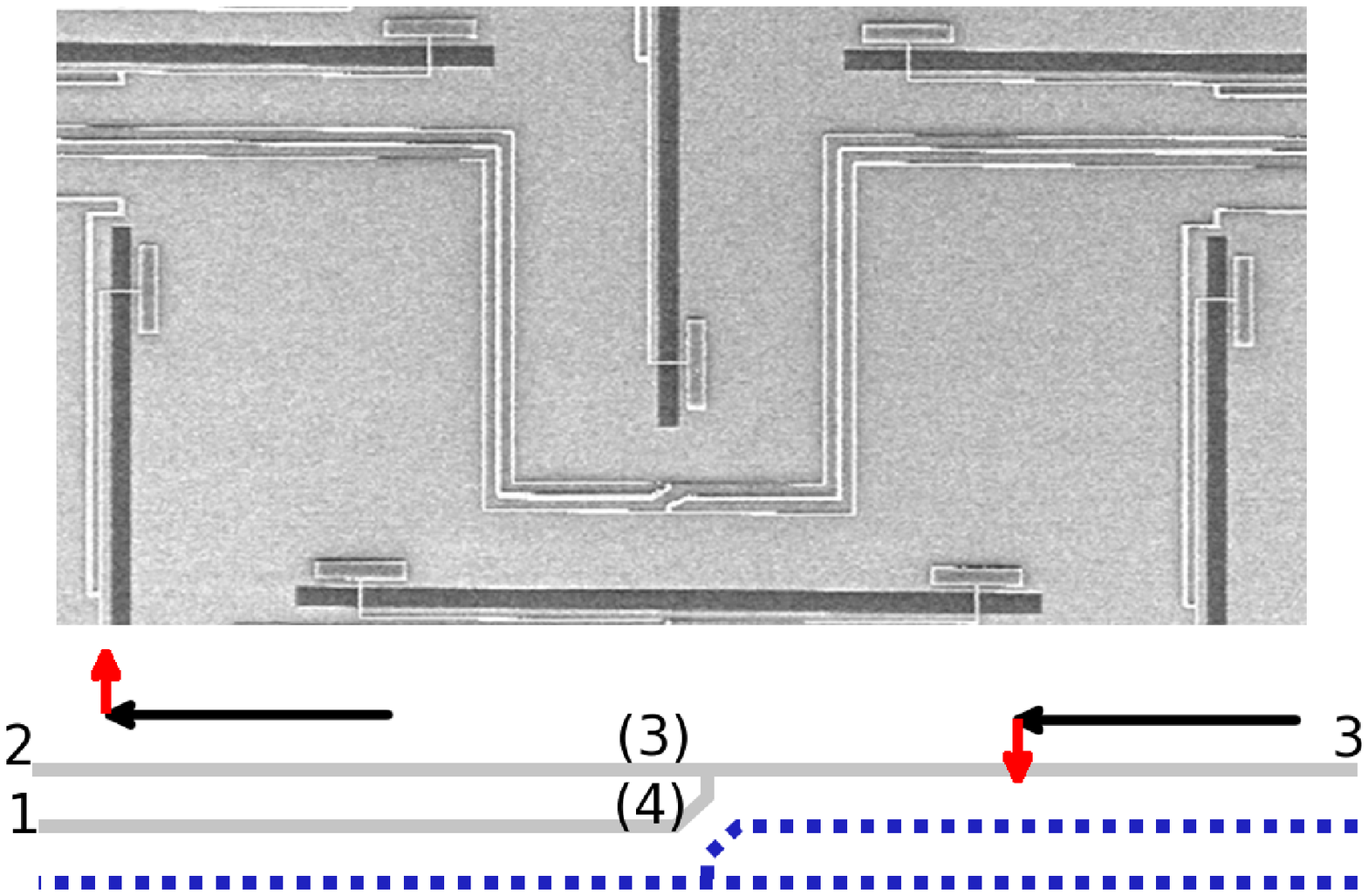}
	\label{mid_tree_pic}
	}
\caption{Focal Plane and Detector Design.  Fig. \ref{FPU_pic} shows a whole focal-plane with four tiles while Fig \ref{pixel_pic} shows one pixel.  The pixel footprint is dominated by the antenna and feed network, and Fig \ref{mid_tree_pic} shows the center of that network where microstrip cross-talk can degrade the beam synthesis.\label{FPU_detectors}}
\end{center}
\end{figure}

Figure \ref{FPU_pic} shows a a BICEP2 focal plane; those for the Keck Array are nearly identical, while those for SPIDER employ modifications for enhanced magnetic shielding.  Each contains four tiles, each of which have a monolithic imaging array with 64 pixels.  We couple optical power onto each pixel (Figure \ref{pixel_pic}) with a planar dual-polarized phased-array antenna\cite{Kuo_antenna_coupled_TES}.  All antennas deployed to date use a 12x12 array of subradiators, where each subradiator contains echelon pairs of 412 $\mu$m x 12 $\mu$m slots etched into a 0.15 $\mu$m thick superconducting niobium (Nb) ground plane.  The opposite polarization's slot pairs interlock with these with a common center, such that the two polarizations' antennas in each pixel are co-located.  Each pixel's antenna is 7.2 mm on  side, providing bare beams (i.e. without any gain from optics) with 14$^o$ FWHM that nicely match the f/2.1 camera optics.

Each slot couples to a microstrip summing network at a pair of feedpoints placed close enough to the slot ends to maintain a low impedance of 47+j15 $\Omega$.  We tune away the reactance with shunt capacitors and match the radiation resistance with 0.93 $\mu$m wide microstrip.  This upper layer is also Nb that is 0.4$\mu$m thick and separated from the ground by 0.3 $\mu$m of sputtered silicon dioxide (SiO2).  The summing network combines the waves from the slots in a series of microstrip-tees, first summing slots across different columns of each rows, then summing the total waves in each row.  Each polarization has its own independent summing tree whose horizontal arms must interlock with the other polarization.

After all the slots' waves are summed onto one microstrip line per polarization, we pass each through a band-defining 150GHz filter with 25\% bandwidth.  The filters have  three-poles with three short stretches of CPW acting as series inductors that are separated by T-networks of capacitors.  Neither the summing tree nor filter require any microstrip vias or microstrip-cross-overs.

We terminate power in a lossy meandered gold microstrip line in close thermal contact with the bolometer's TES.  The gold is deposited before the Nb and cleaned with an Ar ion mill to provide a reliable electrical contact with the superconducing microstrip.  The meander's length is 2.2mm, which provides a return loss of less than -20dB.  We pattern two TESs in series on each bolometer made of Aluminum (Al) and Titanium (Ti) which respectively have nominal superconducting transition temperatures $T_c$ of 1.2K and 0.5K.  We voltage bias into the Al transition for lab testing where we need to expose the camera to 300K and into the Ti transition for science observations, where the effective sky temperature is much lower.  The bolometer itself is suspended on a 1$\mu$m thick film of low stress nitride released by a XeF$_2$ etch.  We select the legs' length and geometry to control the thermal conductance (G), and hence the thermal carrier noise and saturation power.

We read the current through the bolometers with a SQUID-based time domain multiplexer system developed by NIST and UBC.  Multiplexing chips and Nyquist filter chips are visible on the perimeter of the focal planes in Figure \ref{FPU_pic}.

\section{DIFFERENTIAL POINTING OF POLARIZATION PAIRS}
\subsection{BEAM SYSTEMATICS}

Improving on BICEP-1's constraint of $r<0.72$\cite{BICEP1_Emodes} requires large imaging arrays of detectors to attain higher sensitivity.  Additionally, the two cross-polarized detectors in each pixel should view the sky through the same optics path; this way, differencing within each detector pair should greatly reduce sky noise.  While our monolithic and planar arrays of dual-polarized bolometers have provided an expedient way to meet these needs, we have a significant burden to synthesize useable beams with the phased-array antennas.  

\begin{figure}[ht]
	\centering
		\includegraphics[scale=0.3]{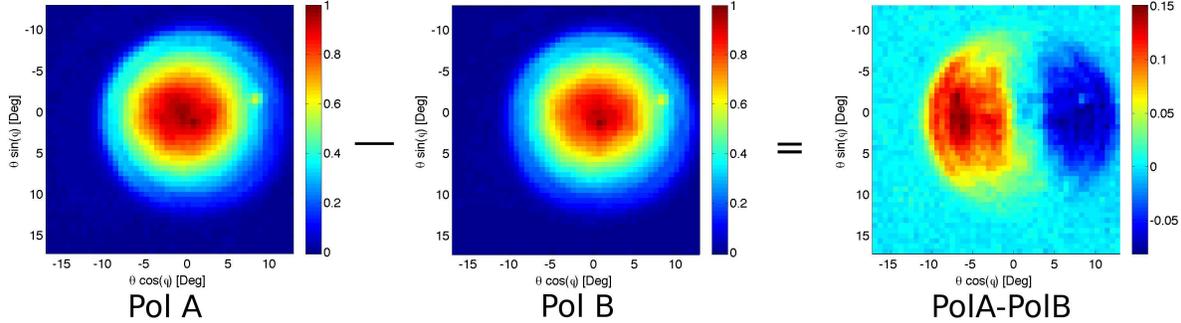}
	\caption{Polarization Pair and difference from BICEP2 illustrating the dipole structure in near field maps.  {\bf We have since reduced the peak-to-peak amplitude to from $\pm7$\% of peak power to $\pm1$\%}}
	\label{dipole_pic}
\end{figure}

Ideally, beam patterns of the two polarizations in a pair would subtract perfectly.  In practice, this subtraction often leaves residual structure that can leak temperature anisotropies into the polarization channels\cite{Shimon_systematics}.  Between beams with a Gaussian profile, (\textbf{a}) differential widths can generate a monopole difference pattern, (\textbf{b}) differential pointing of centroids can generate a dipole difference pattern, and (\textbf{c}) differential ellipticity can generate a quadrupole difference pattern.  Of these defects, the dipoles (\textbf{b}) are by far the most dominant (See Figure \ref{dipole_pic}), with an angular separation between centroids as high as 10\% of the FWHM as measured in the far field.

Dipoles can leak the much brighter temperature gradients into the polarization channels.  BICEP2 has mitigated this problem by co-adding maps taken at different boresight rotations and they are currently developing an analysis technique to remove leaked B-modes by regression against temperature maps.  Additionally, Spider will see little sky noise, obviating their need to difference the polarization pairs' time-streams.  Nonetheless, the BICEP-1 team concluded that for their scan strategy and analysis pipeline at the time, they would need differential pointing less than 1.9\% to avoid a false B-mode signal of $r=0.1$, and we have taken this figure as a benchmark for our detector design.

\subsection{HORIZONTAL DIPOLES}

BICEP2 has seen differential steering of 7\% of the beam width in the camera's near field (i.e. near the aperture) on the axis parallel to the antenna's summing tree rows (horizontal), but very little in the direction along columns (vertical).  They have measured comparably large dipoles in the far field, although with an orientation that varies between detectors and not strictly horizontal \cite{Aiken_SPIE}.  While the offsets in the near-field should not generate offsets in the far-field, an optical misalignment that moves the beam aperture locations from the location of the stop could allow such leakage.  We have not yet succeeded in finding a misalignment, but in an effort to make our cameras robust against such alignments errors, we have redesigned the detectors to remove the near-field dipole.

The near-field dipoles from BICEP2 always maintain the same horizontal orientation in every tile with a very narrow spread in magnitude, which suggested a problem related to the microstrip artwork.  Cross talk between adjacent parallel microstrip lines can in fact generate horizontal differential steering.  Figure \ref{mid_tree_pic} shows the interlocking horizontal summing trees from both polarizations, focusing on one of the few places that our design breaks left-right symmetry.  The lines that connect the end branches of the vertical trees to the top of the horizontal trees run parallel to top branch of the horizontal trees for a length of 3mm, or 3.75 wavelengths at band-center.  In early designs, the lines had an average separation of 8 $\mu$m, center-to-center.  

These microstrip pairs effectively form a forward-wave directional coupler, but only on the half of the horizontal tree closest to the vertical tree.  The fields of the odd mode on the coupled lines (same current, opposite phase) fringe more out of the silicon oxide than those of the even mode (same current, same phase), allowing the modes to propagate with different wave speeds.  In reference to the port labels in Figure \ref{mid_tree_pic}, a wave launched from the vertical tree towards the horizontal (from port 1) is a superposition of these even and odd modes that cancel in the adjacent line at the end nearest the vertical tree (Port 2).  However, the modes' different wave speeds guarantee that they will not cancel in the adjacent line at the end nearest the pixel center (Port (3)). We label ports (3) and (4) in parenthesis because they are subsequently cascaded in the microstrip tee, thus becoming internal ports.  If we explicitly sum the scattering parameters\cite{Coupled_lines_book},

\begin{eqnarray}
 S_{(4)1}=(S^e_{(4)1}+S^o_{(4)1})/2=(e^{-jk_e\ell}+e^{-jk_o\ell})/2 &=& \quad e^{-j(k_e+k_o) \ell/2}\cos ((k_e-k_o)\ell/2)      \nonumber \\
 S_{(3)1}=(S^e_{(3)1}+S^o_{(3)1})/2=(e^{-jk_e\ell}-e^{-jk_o\ell})/2  &=& -je^{-j(k_e+k_o) \ell/2}\sin ((k_e-k_o)\ell/2)
\end{eqnarray}
\\*then we see that this coupled wave is small in amplitude and lags the through wave (in Port (4)) by 90$^o$.  We split the through wave with even power in a microstrip tee, but the coupled wave adds in quadrature to waves on the side opposite the vertical tree, retarding the phase.  This phase step steers the two polarizations' beams away from the vertical tree, thus generating differential pointing in the horizontal direction.

\begin{figure}[ht]
	\centering
		\includegraphics[scale=0.5]{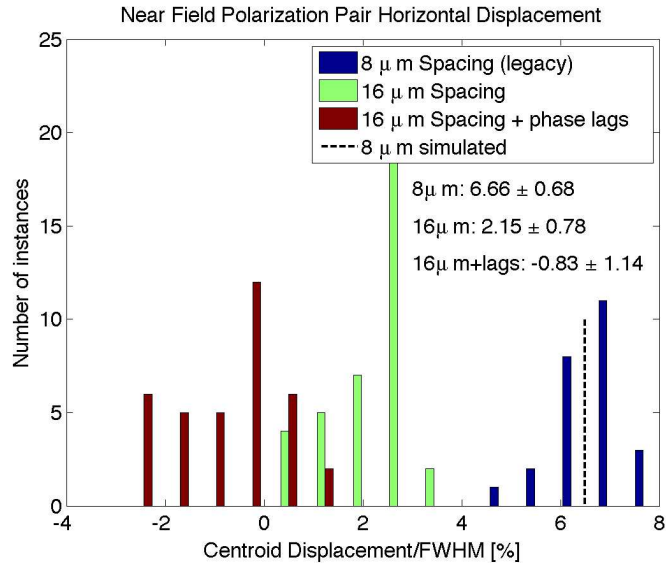}
	\caption{Polarization Pair Centroid displacement normalized to FWHM in the near field.  We show histograms of the tiles with two different spacings as well as one where built phase lags into the summing tree to manually tune this to zero.}
	\label{histo_nfbm}
\end{figure}

Simulations using a variety of commercial packages (HFSS, Sonnet, ADS Momentum) show that the coupled wave amplitude is a function of dielectric permittivity and thickness, the spacing between the lines, and the thickness of the metal conductors.  The upper Nb film thickness of 0.4 $\mu$m exceeds that of the dielectric (0.3 $\mu$m) and is a substantial fraction of the line separation, so it can contribute to the capacitive coupling in a substantial way. If we neglect this thickness, then we underestimate the coupling by a factor of 40\%.   Accounting for the Nb microstrip thickness in simulation provides horizontal differential steering that agrees with measurements as shown in the black dashed line in Figure \ref{histo_nfbm}.  This understanding has proven useful for the SuperSpec team that uses microstrip couplers to drive their spectrometer resonators\cite{Shiro_SPIE_Superspec}.

The spacing between horizontal lines is constrained by the need to route between slots in the tree.  Nonetheless, we have managed to rearrange the summing tree in ways that increase the average separation between lines.  Figure \ref{histo_nfbm} shows the splitting from several data tiles with different average spacing, and we have reduced the steering to less than 2\% of FWHM\cite{LTD14_OBrient}.  Finally, we have fabricated tiles with residual microstrip coupling but have tuned the steering to zero by introducing intentional phase lags (extra line length) on the sides furthest from the vertical trees.  Figure \ref{histo_nfbm} also shows that those have a steering reduced to an average within a standard deviation of zero on the horizontal axis.  We are currently fabricating  detectors with summing trees that have an average separation of 50 $\mu$m between the top lines of the horizontal tree; we expect this to have no measurable horizontal offset, which will obviate the phase-lag correction.

\subsection{VERTICAL DIPOLES}

The Keck Array cameras have shown evidence of vertical near field components in addition to the horizontal ones.  Unlike the horizontal components described above, these have a large scatter in magnitude, although a constant orientation.  This large variance suggested that the problem was not due to the microstrip artwork but was probably related to non-uniform film properties.

Between the BICEP2 and Keck Array deployments, we switched to a Nb sputter system with a larger gun to attain better uniformity.  However, this system ran at significantly higher power than the old one and lacked grounding to shunt away higher energy particles.  Until recently, we have used liftoff to define our microstrip artwork, and we now believe that the elevated temperatures in the sputter system allowed contaminants to leach out of the photoresist and into the Nb films during sputtering.  RRR measurements of lines defined by etching exceed those defined by liftoff by nearly 40\%.  In a few extreme cases, we have seen visible blackening of the lifted-off microstrip lines.

A superconductor is considered dirty when the mean-free path between scattering from impurities $\ell$ is much less than the London depth $\lambda_L$ or the Cooper-pair coherence length $\xi_o$.  In this limit, the effective penetration depth is \cite{Tinkham_book}

\begin{equation}
\lambda_{eff}=\lambda_L\sqrt{\frac{\xi_o}{\ell}}
\label{eqn_pen}
\end{equation}

Niobium has $\lambda_L=52$nm $\xi_o=$39nm\cite{Poole_book}.  We have fabricated test devices that include stretches of transmission line mismatched in impedance from the surrounding lines\cite{Kuo_antenna_coupled_TES}.  These form Fabrey-Perot cavities whose standing wave period is sensitive to the niobium's penetration depth.  Our films appear to have $\lambda_{eff}$=110nm, which is similar to measurements of other researchers in the field\cite{Kerr_ALMAmemo}.  Equation \ref{eqn_pen} suggests that the mean free path between scatterings is $\ell=10$nm, which puts us well into the dirty limit in which film cleanliness can impact the circuit performance.

Non-uniform contamination can produce non-uniform kinetic inductance, which in turn can spatially perturb the wavespeeds in the microstrip summing tree.  Variations in wavespeed can steer beams off boresight.  Additionally, the summing tree does not treat the two polarizations identically, and as a result, they can steer differentially.  We have observed that the tiles with the largest scatter in beam centroid position correspond to those with the largest vertical dipole components.  We also expect the tree to induce larger steering in the vertical than horizontal because slots along rows combine immediately in the horizontal tree resulting in less integrated phase error than those along columns that combine in the vertical tree after horizontal summing.

To understand the effects of nonuniform niobium contamination, we have written a circuit model of our summing tree.  We use a quasi-static microstrip model\cite{Hammerstad_Jensen} for the lines and simple power division models for the microstrip tees.  The code uses user-provided arrays of film properties as a function of physical position to locally compute the impedances and wavespeeds for differential (1$\mu$m long) line segments.  If we cascade two sequential ($m+\ell$) port and ($\ell+n$) port circuits with scattering matrices $S^1$ and $S^2$ respectively through common ports $\ell$, then the combined circuit matrix is\cite{S_matrix_cascade}:

\begin{equation}
S^{1+2}=
 \begin{bmatrix}
  S^1_{m,m}+ S^1_{m,\ell}S^2_{\ell,\ell}\left(I_{\ell,\ell}-S^1_{\ell,\ell}S^2_{\ell,\ell}\right)^{-1}S^1_{\ell,m}& S^1_{m, \ell}\left( I_{\ell,\ell}-S^2_{\ell, \ell} S^1_{\ell, \ell} \right)^{-1}S^2_{\ell, n}  \\
  S^1_{m, \ell}\left( I_{\ell,\ell}-S^2_{\ell, \ell} S^1_{\ell, \ell} \right)^{-1}S^2_{n, \ell}  &   S^2_{n,n}+ S^2_{n,\ell}S^2_{n,n}\left(I_{\ell,\ell}-S^2_{\ell,\ell}S^1_{\ell,\ell}\right)^{-1}S^2_{\ell,n} 
 \end{bmatrix}
\end{equation}

\begin{figure}[ht]
	\centering
		\includegraphics[scale=0.5]{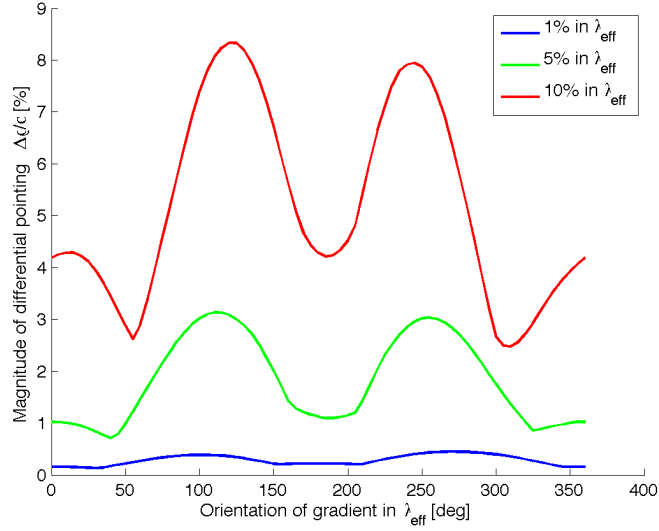}
	\caption{Simulated Magnitude of differential pointing as a function of orientation of $\lambda_{eff}$ gradient orientation for different gradient magnitudes .  The steering is a minimum near angles 45$^o$ from the polarization planes, where the gradient effects both polarizations equally and it is twice as large for the vertical as horizontal.  The magnitude is over twice as large when oriented on the vertical axis than horizontal}
	\label{gradient_plots}
\end{figure}

We implemented this algorithm in software that is functionally similar to commercial packages like Agilent's ADS circuit simulator, except that it includes spatially varying film properties.  Figure \ref{gradient_plots} shows the computed dipoles that arise from linear gradients in the penetration depth $\lambda_{eff}$ versus orientation of that dipole.  Each curve corresponds to a different film gradient magnitude, and a gradient of 5\% in penetration depth across a pixel can generate measured vertical dipoles similar to our measured levels.  In reality, the spatial variations of $\lambda_{eff}$ may be more complicated than this, but these computations demonstrate that, to first order, differential steering can arise from film gradients.

\begin{figure}[ht]
\centering
\subfigure[Near-field centroid displacement of \emph{liftoff} niobium processed tile \label{Liftoff_NFhist}]{
\includegraphics[scale=0.4]{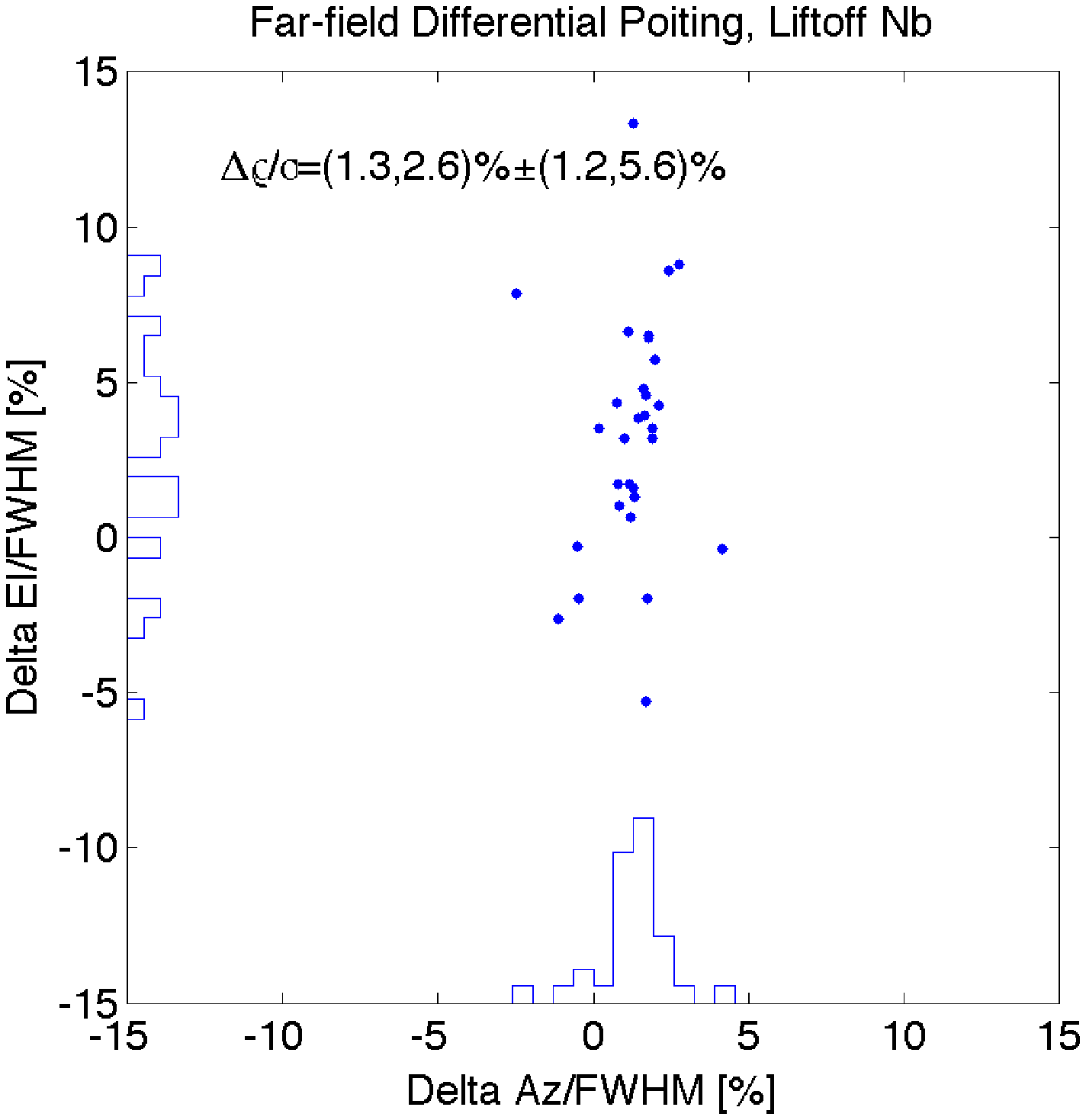}
}
\subfigure[Near-field centroid displacement of \emph{etchback} niobium processed tile \label{etched_NFhist}]{
  \includegraphics[scale=0.4]{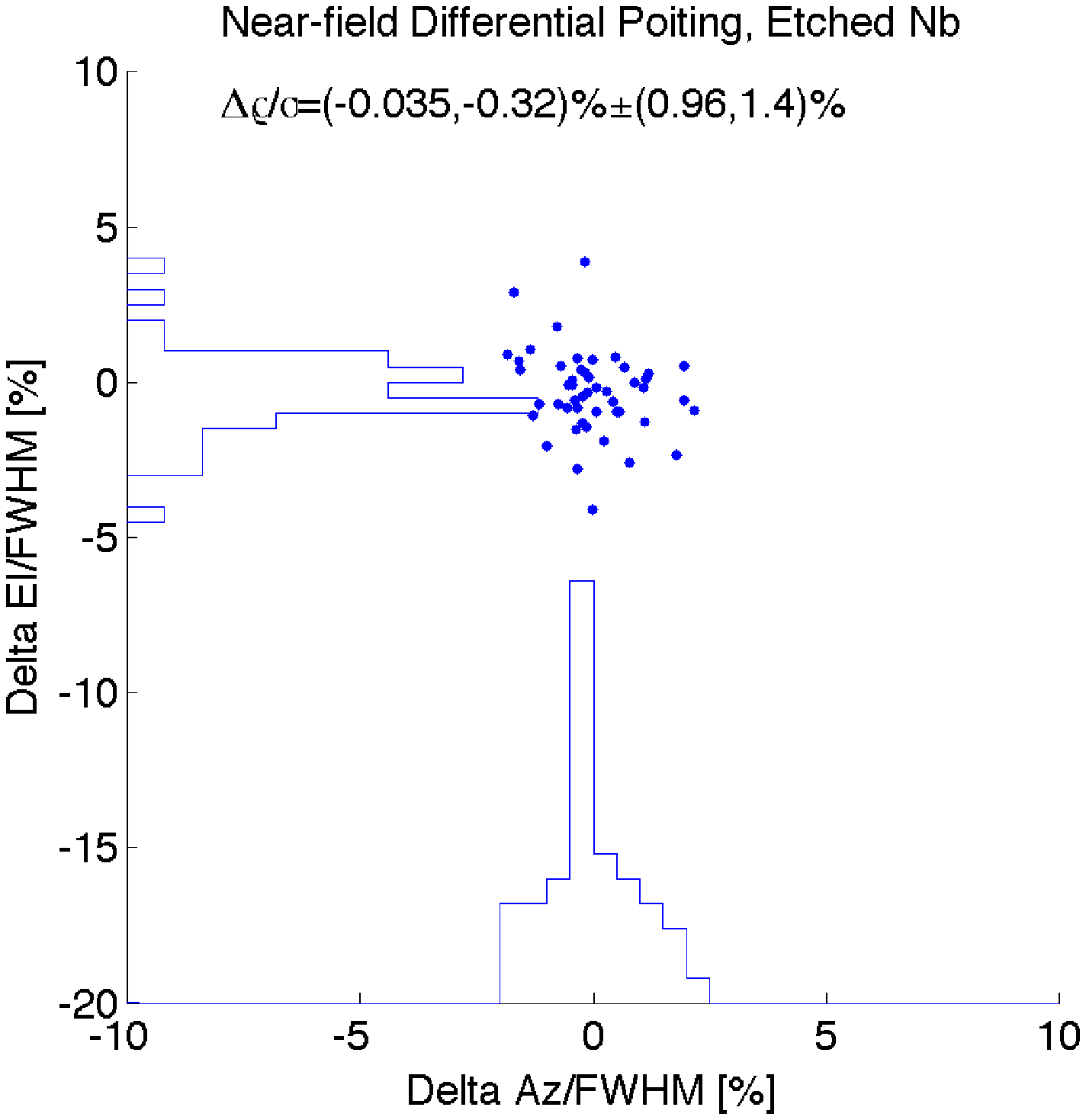}
}
\subfigure[Far-field centroid displacement of \emph{liftoff} niobium processed tile \label{Liftoff_FFhist}]{
\includegraphics[scale=0.4]{Liftoff_FFmap.eps}
}
\subfigure[Far-field centroid displacement of \emph{etchback} niobium processed tile \label{etched_FFhist}]{
  \includegraphics[scale=0.4]{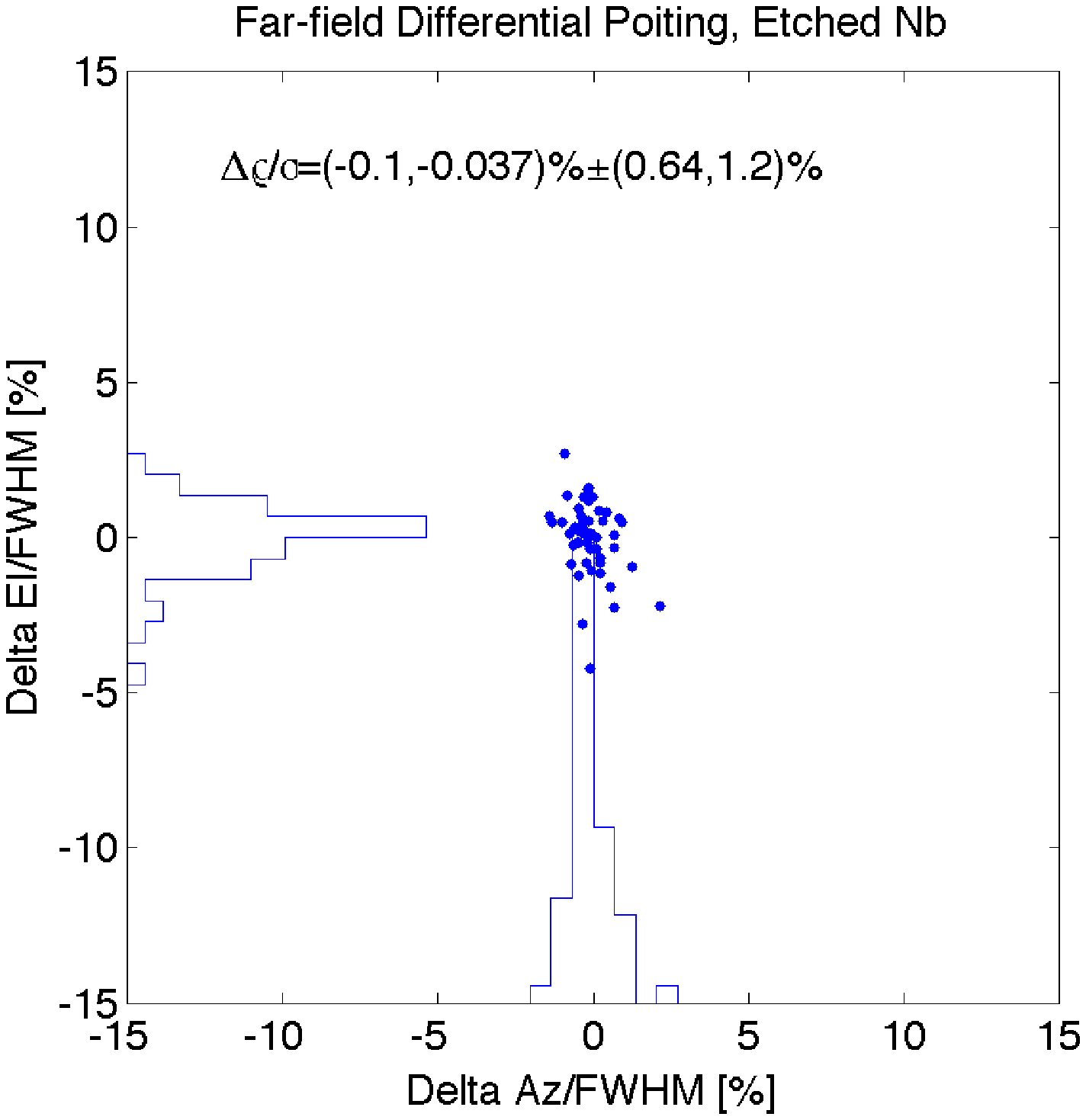}
}
\caption{Polarization Pair Centroid displacement measured in the near and far fields of a Spider test-bed camera, all measured during the same cryogenic run.  We plot the displacements as a scatter-plot, but also show histograms of the two dimensions at the edges.  The two tiles have different processing in the microstrip niobium, but are otherwise identical, demonstrating that the niobium processing can have a dramatic effect on beam synthesis.  Two other tile pairs from the same run show similar results, as well as another four tiles in a test-bed without optics.\label{2D_histograms}}
\end{figure}

We have changed our niobium processing from lift-off to etching to reduce this contamination.  This simple change has created a dramatic improvement in polarization-pair centroid alignment.  Figure \ref{2D_histograms} shows displacements of best fit centroids in both the near and far-fields of a Spider testbed.  At this point, the alignment is better than the BICEP-1 benchmark.

\section{SIDELOBE SUPRESSION}
As mentioned above, BICEP2 and Keck Array cameras all have stops at their apertures cooled to 4K (Figure \ref{B2_rays}); the Spider team cools theirs to 2K.  This has let the three projects use the simplest antenna design where each slot provides equal power to the bolometer.  The summing tree for this antenna is relatively easy to design, but the uniform (tophat) illumination generates a sinc-function pattern with side lobes at -13dB of peak power.  These side lobes terminate on the cold stop in these cameras.  

However, large throughput crossed-Dragone telescopes like Polar-1 shown in Figure \ref{P1_rays} have an aperture beyond the large primary where it is difficult to be made cold.  In lieu of a traditional absorbing cold stop, Polar-1 will use an aluminum Gaussian scattering surface and a larger Winston cone to direct the spillover onto the sky.  This will create an effective stop with a temperature of 15K.  

In parallel to this novel optical design, we have redesigned the detectors to reduce spillover.  We do this by illuminating the subradiator slots with a Gaussian illumination pattern, which synthesizes a beam with a lower sidelobe level, albeit with a wider main lobe \cite{Dolph_array}.  We can ultimately construct an arbitrary illumination pattern by choosing the line widths, and hence impedances, at each of the mictostrip tees in the feed network.  In the uniform illumination tree, most junctions split power evenly (in a time reversed picture).  However, most of the junctions in the Gaussian illuminated tree split power asymmetrically.

Only a subset of the the subradiating slots are on the sky-side of a given tee-junction.  This subset further partitions into ``inside'' slots connected to the microstrip line directed to the antenna's interior, and ``outside'' slots connected to the microstrip line directed to the antenna's exterior.  The required power division at each junction is:

\begin{equation}
P_{inside}=\left( 1-\frac{\displaystyle \sum_{outside}P_{slot}}{\displaystyle \sum_{inside}P_{slot}}\right)^{-1} 
\label{taper_design}
\end{equation}
\\*and $P_{outside}=1-P_{inside}$.  The ratio of impedances $Z_{inside}/Z_{outside}$ on the slot side lines must be the reciprocal of the power ratio i.e. $P_{outside}/P_{inside}$, subject to the requirement that the impedance be matched when looking into the junction from the bolometer side.  Once the desired distribution of powers in the slots has been chosen, we use these formulae to design each of the microstrip tees in the feed network.

We have optimized our feed to minimize spillover onto Polar's stop of f/1.52.  For pixels that are 7.2mm on a side (12x12 subradiators), the spillover can be as low as 2.6\%, while for a more ambitious 6mm pixel (10x10 subradiators), the spillover should be 6.2\%.  In both cases, the optimal waist of the Gaussian feed pattern is 3.6mm, but in the 6mm pixel case, the beam truncates more abruptly, generating larger sidelobes.

\begin{figure}[ht]
\centering
	\includegraphics[scale=0.4]{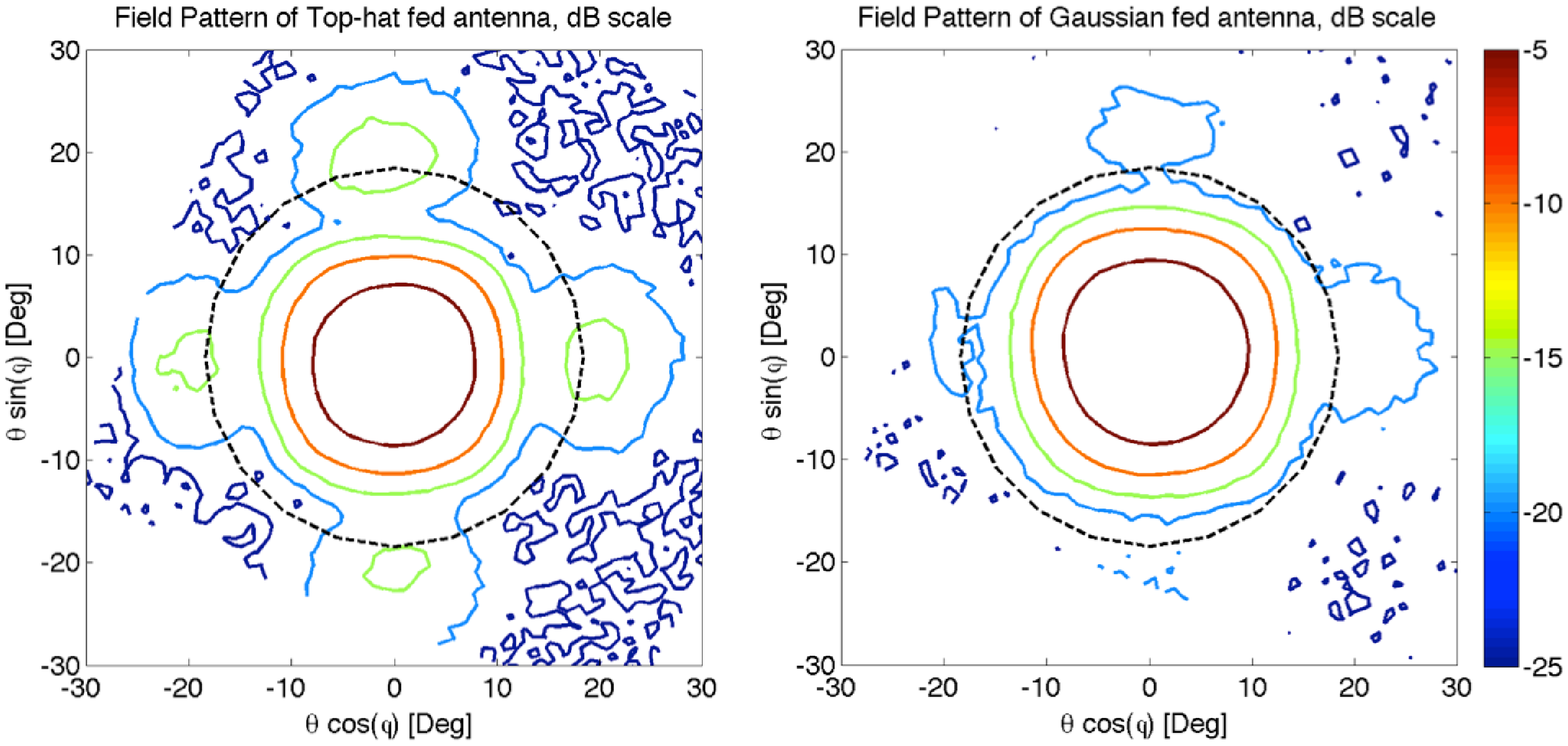}
	\caption{Beam Patterns of tophat and Gaussian illuminated antennas, where Polar-1's stop is shown in the dashed circle. \label{tapered_pattern}}
\end{figure}

We fabricated several test prototypes pixels with an illumination waist of 4.4mm and measured their beam patterns in a test cryostat without refracting optics or a stop.  Figure \ref{tapered_pattern} shows measurements of a pair of patterns taken from a tophat and Gaussian illumination, both 7.2mm on a side.  In this case, the sidelobes are supressed from -12dB to -17dB while the main lobes have increased from 14$^o$ FWHM to 16.5$^o$ FWHM.  For this tile, we found that the tophat antennas had 47\% total receiver efficiency while the guassian antennas had 49\% \cite{LTD14_OBrient}.  This confirms that the tapered illumination is generated by receiving less power at the edge, but more in the center, rather than less power overall.

\begin{figure}[ht]
	\centering
		\includegraphics[scale=0.4]{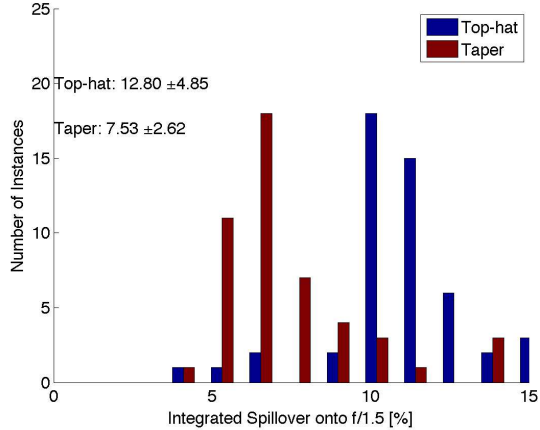}
	\caption{Spillover histograms for tophat and spillover antennas, each 7.2mm on a side (12.x12 format)}
	\label{spillover_plot}
\end{figure}

Using the measured pattern waist and sidelobe level, the total spillover onto f/1.52 from these prototype pixels should be 7.1\%.  We can also directly integrate the total spillover beyond the stop ring shown in Figure \ref{tapered_pattern}, and we find 7.5$\pm$ 3.0 \%, which agrees with the figure computed from the beam width and side lobe levels.  Note that the tophat would spill 12.8\% (see Figure \ref{spillover_plot}).

We have since fabricated devices with the optimal waist sizes for three different pixel sizes: 7.2mm (12x12), 6.0mm (10x10), and 4.8mm (8x8).  Due to a fabrication error, the efficiency of that batch was low, leading to a low signal to noise measurement and a diffuse response of -17dB per map pixel that we suspect is from direct stimulation of the bolometer.  In a focal with a higher efficiency, this direct stimulation is below -25dB per map pixel.  As a result, it is difficult to characterize the side-lobe level in these devices.

\begin{figure}[ht]
	\centering
		\includegraphics[scale=0.4]{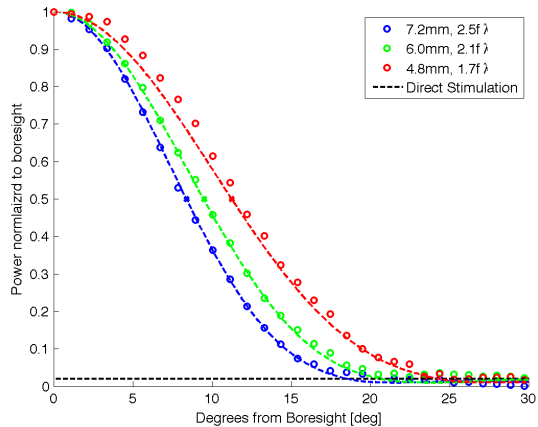}
	\caption{E-plane cuts for the three different sized pixels.  The colored dashed lines are the expected beams from simulation and the crosses are the half-power points interpolated from the measurement}
	\label{cuts_plot}
\end{figure}

However, the measured beam widths of 16.8$^o$, 19.0$^o$, and 22.4$^o$ FWHM match the expected widths, so we anticipate that they will have spillovers of roughly 2.5\%, 6.2\%, and 14\%.  Figure \ref{cuts_plot} shows cuts from 36 stacked maps on a linear scale to emphasize the FWHM of the main lobe, and co-plots the model in dashed lines to emphasize the agreement.  We also show the direct stimulation floor to emphasize how we cannot really characterize the side-lobes aside from placing an upper bound of -17dB.

\section{Future Work}
We have demonstrated differential steering that is below the 1\% of FWHM standard set by the BICEP-1 team.  We will upgrade detectors from the Keck Array this winter, and focal planes being made now for Spider's 2013 flight have these fixes.  We will refabricate the detectors with the tapered illumination to check the side lobe levels before we start populating Polar-1's focal plane this fall.

\acknowledgments 
 
R. O'Brient and Z. Staniszewski would like to thank Oak Ridge Associated Universities for funding through the NASA Post-doctoral Program.  Keck Array is partially funded by the Keck Foundation and the detector development through the Betty and Gordon Moore Foundation.  All devices were fabricated in the Microdevices Laboratory (MDL) at NASA's Jet Propulsion Laboratory.

\bibliography{SPIE_bib}   
\bibliographystyle{spiebib}   

\end{document}